\documentclass[prl,twocolumn,showpacs]{revtex4}

\usepackage[german,english]{babel}
\usepackage{amssymb}
\usepackage{amsfonts}
\usepackage{amsmath}
\usepackage{mathrsfs}
\usepackage{graphicx}

\begin{document}

\title{Coupled dynamics of DNA-breathing and single-stranded DNA 
binding proteins}

\author{Tobias Ambj{\"o}rnsson}
\email{ambjorn@nordita.dk}
\affiliation{NORDITA, Blegdamsvej
17, DK-2100 Copenhagen {\O}, Denmark}
\author{Ralf Metzler}
\email{metz@nordita.dk}
\affiliation{NORDITA, Blegdamsvej
17, DK-2100 Copenhagen {\O}, Denmark}



\begin{abstract}
We study the size fluctuations of a local denaturation zone in a DNA molecule
in the presence of proteins that selectively bind to single-stranded DNA, based
on a $(2+1)$-dimensional master equation.
By tuning the physical parameters we can drive
the system from undisturbed bubble fluctuations to full, binding
protein-induced denaturation. We determine the effective free energy
landscape of the DNA-bubble and explore its relaxation modes.
\end{abstract}

\maketitle

Under physiological conditions the Watson-Crick double-helix is the
thermodynamically stable configuration of a DNA molecule. This stability
is effected by the specific Watson-Crick H-bonding, whose key-lock principle
guarantees the high level of fidelity during replication and transcription;
and by the stronger base-stacking between neighboring base-pairs causing
hydrophobic interactions between the planar aromatic bases
\cite{kornberg,delcourt}.

The initial breaking of the stacking interactions in an unperturbed DNA
molecule is associated with an activation barrier $\sigma_0\simeq 10^{-3
\ldots -5}$ \cite{wartell,blake,blossey}. Once this barrier is overcome, the
free energy of breaking an additional base-pair
is of the order of 1 to 2$k_BT$ \cite{santalucia,blake}, effecting
local, single-stranded DNA denaturation zones of a few tens
of broken base-pairs \cite{gueron}. By thermal activation, these 
\emph{DNA-bubbles} fluctuate in size by a zipper motion at the
two forks where the bubble meets the intact double-strand
\cite{kittel}. By fluorescence correlation methods
this \emph{DNA-breathing} can be explored on the single molecule level,
revealing multistate (un)zipping kinetics with a typical single (un)zipping
step time scale around 50$\mu$s and a typical bubble lifetime of the order of
10 ms \cite{altan}.

A long standing puzzle had been why the presence of selectively single-stranded
DNA binding proteins (SSBs) does not lead to full DNA-denaturation, as SSB
binding is thermodynamically favorable \cite{jensen}. Detailed single-molecule
optical tweezers studies, by overstretching of the DNA molecule to bring the
effective temperature close to the melting temperature $T_m$ in the presence
of bacteriophage T4 gene 32 SSBs, showed quantitatively \cite{pant} that there
exists a kinetic block for SSB-binding \cite{karpel}: below $T_m$ the bubble
lifetime is shorter than the typical SSB binding time, counteracting
helix-destabilization.

\begin{figure}
\includegraphics[width=6.8cm]{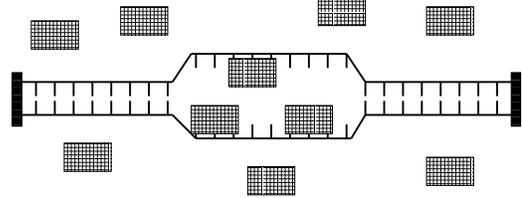}
\caption{Clamped DNA bubble in a region of size $M$,
immersed in a bath of SSBs. SSBs do not bind across
zippers.}
\label{fig1}
\end{figure}

In what follows, we develop a dynamical model to quantify the coupled dynamics
between a fluctuating DNA-bubble and SSBs that attempt to bind to it, in terms
of the system parameters (temperature, external force, SSB binding rate and
strength, SSB size). We demonstrate that the presence of SSBs leads to
enhanced bubble lifetime. Effectively, the bubble free energy is lowered,
and even SSB-induced denaturation can occur.

The system we have in mind (Fig.~\ref{fig1}) resembles the DNA-construct
from Ref.~\cite{altan}, in which a homopolymer bubble region is clamped
at both ends. The one-bubble approximation used here is generally
valid below $T_m$ due to $\sigma_0\ll 1$. Fig.~\ref{fig1} also illustrates that
the typical binding size $\lambda$ of an SSB
($\simeq 10$ bases), is of the same order as the bubble size. It
is therefore necessary to consider the statistical weight from SSBs explicitly,
instead of defining an effective chemical potential \cite{ambme}.

\begin{figure}
\includegraphics[width=8cm]{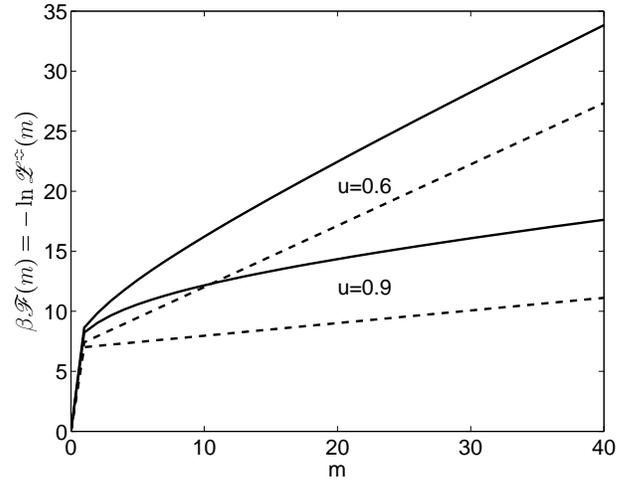}
\caption{Bubble free energy in absence of SSBs as function of bubble
size $m$ (---:~$c=1.76$; - - -:~$c=0$). We chose $\sigma_0=10^{-3}$.}
\label{fig2}
\end{figure}

To quantify the system, we define the probability distribution $P(m,n,t)$
to find a bubble of size $m$, with $n$ bound SSBs at time $t$.
Its time evolution is governed by the (2+1)-dimensional master equation
\begin{widetext}
\begin{eqnarray}
\nonumber
\partial P(m,n,t)/\partial t&=&\mathsf{t}^+(m-1,n)P(m-1,n,t)+
\mathsf{t}^-(m+1,n)P(m+1,n,t)\\
\nonumber
&&-\left( \mathsf{t}^+(m,n)+\mathsf{t}^-(m,n)\right) P(m,n,t)\\
\nonumber
&&+\mathsf{r}^+(m,n-1)P(m,n-1,t)+ \mathsf{r}^-(m,n+1)P(m,n+1,t)\\
&&-\left( \mathsf{r}^+(m,n)+\mathsf{r}^-(m,n)\right) P(m,n,t),
\label{master}
\end{eqnarray}
\end{widetext}
owing to the discrete nature of the problem. In Eq.~(\ref{master}),
the transfer rates $\mathsf{t}^{\pm}$ describe changes in
the bubbles size $m$, and $\mathsf{r}^{\pm}$ changes in the number $n$ of
bound SSBs. The boundary conditions for the $\mathsf{r}^{\pm}$ are reflecting
at $n=0$ and $n_{\mathrm{max}}=2[m/\lambda]$, the maximum number of SSBs that
can bind to a bubble of $m$ denatured base-pairs. Similarly,
we impose a reflecting boundary condition at $m=0$ and $m=M$, the maximum
bubble size. Moreover, we have to consider that a bubble cannot zip close if
its size is a multiple of the SSB-size, $m=k\lambda$ ($k\in\mathbb{N}$) and
the number of bound SSBs is $n_{\mathrm{max}}$ or $n_{\mathrm{max}}-1$, i.e.,
at least one of the arches of the bubble is fully occupied. To define the
transfer rates based on the statistical weight $\mathscr{Z}(m,n)$
to find a state $(m,n)$, we assume detailed balance \cite{ambme1,risken}:
\begin{eqnarray}
&&\mathsf{t}^+(m-1,n)\mathscr{Z}(m-1,n)=\mathsf{t}^-(m,n)\mathscr{Z}(m,n)\\
&&\mathsf{r}^+(m,n-1)\mathscr{Z}(m,n-1)=\mathsf{r}^-(m,n)\mathscr{Z}(m,n).
\end{eqnarray}
The statistical weight $\mathscr{Z}(m,n)=\mathscr{Z}^{\Bumpeq}(m)\mathscr{Z}
^{\mathrm{SSB}}(m,n)$ has two contributions. The bubble part according to
the Poland-Scheraga model for DNA-melting is \cite{poland,richard}
\begin{equation}
\label{z_bubble}
\mathscr{Z}^{\Bumpeq}(m)=\sigma_0u^m(1+m)^{-c}, \,\, m\ge 1,
\end{equation}
with the bubble initiation factor $\sigma_0$, the weight $u=e^{-\beta E}$ for
breaking a base-pair, and the loop closure factor $(1+m)^{-c}$ for creating
a polymer loop of size $m$, with the offset by 1 due to persistence length
corrections \cite{blake,freire}. We choose typical values, $\sigma_0=10^{-3}$
and $c=1.76$ \cite{richard,REM}.
Eq.~(\ref{z_bubble}) is completed by $\mathscr{Z}^{\Bumpeq}(0)=1$.
In comparison to the continuum description of DNA-breathing in absence of
SSBs by a Fokker-Planck equation involving the gradient of the free
energy \cite{hame}, we note that the discrete approach (\ref{master}) explicitly
includes the constant activation barrier $\sigma_0$, compare Fig.~\ref{fig2}.

The contribution from the SSBs has the form \cite{ambme,ambme1}
\begin{equation}
\mathscr{Z}^{\mathrm{SSB}}(m,n)=\kappa^n\Omega(m,n),
\label{z_ssb}
\end{equation}
with the binding strength $\kappa=c_0K^{\mathrm{eq}}$, involving 
the SSB-concentration $c_0$ and equilibrium binding constant $K^{\mathrm{
eq}}=v_0\exp(\beta|E_{\mathrm{SSB}}|)$, where $v_0$ is the typical SSB volume
and $E_{\mathrm{SSB}}$ its binding energy. The
weight $\Omega(m,n)$ counts all possible ways of putting
$n$ SSBs onto the two arches of the bubble, including potential
gaps between SSBs due to the fact that their size $\lambda$ is larger than a
base. Explicitly,
\begin{equation}
\Omega(m,n)=\sum_{n'=0}^{n}\omega(m,n')\omega(m,n-n')
\end{equation}
with $n,(n-n')\le n^{\mathrm{max}}/2$, and the combinatorial term
\begin{equation}
\omega(m,n)={m-(\lambda-1)n \choose n}.
\end{equation}
By detailed balance and taking the rate $\mathsf{t}^+$ for breaking a
base-pair to be proportional to the activation factor $u$ (including
loop closure effects), and assuming that the rate $\mathsf{r}^-$ for
SSB unbinding is proportional to the number $n$ of bound SSBs, we arrive
at the expression for the transfer coefficients, i.e., the rates for the
bubble size
\begin{subequations}
\begin{eqnarray}
&&\mathsf{t}^-(m,n)=k\Omega(m-1,n)/\Omega(m,n)\\
&&\mathsf{t}^+(m,n)=ku([1+m]/[2+m])^c,\,\, m\ge 1
\end{eqnarray}
\end{subequations}
with the bubble initiation rate $\mathsf{t}^+(0,0)=2^{-c}k\sigma_0u$, and
\begin{subequations}
\begin{eqnarray}
&&\mathsf{r}^-(m,n)=n\gamma k\\
&&\mathsf{r}^+(m,n)=\gamma k\kappa(n+1)\Omega(m,n+1)/\Omega(m,n)
\end{eqnarray}
\end{subequations}
for the SSB number transfer rates \cite{rem}. Here, we introduced the
dimensionless ratio $\gamma\equiv q/k$ of the SSB unbinding rate $q$
and the base pair zipping rate $k$.

\begin{figure}
\includegraphics[width=8cm]{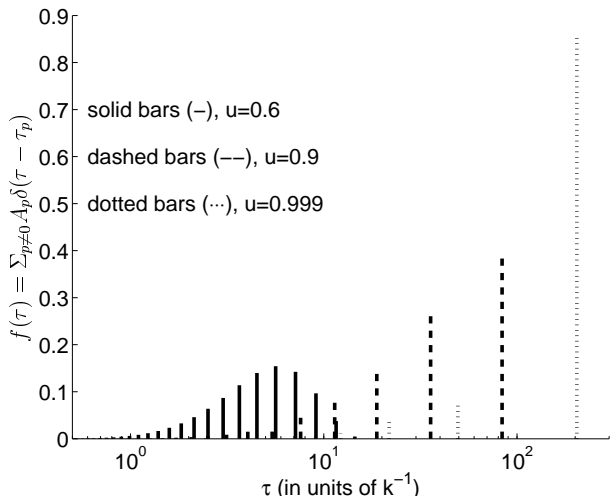}
\caption{Relaxation time spectrum for  $\sigma_0=10^{-3}$
and $M=40$. For $u=0.999$, the longest
relaxation time dominates.
\label{fig3}}
\end{figure}

To solve the master equation (\ref{master}), we introduce an eigenmode
expansion of the form
\begin{equation}
P(m,n,t)=\sum_p c_p Q_p(m,n) \exp\Big( -t/\tau_p\Big),
\label{expansion}
\end{equation}
in which the coefficients $c_p$ of a given eigenmode $p$ are determined
via the initial conditions. The corresponding eigenvalue equation for the
bubble size-SSB number eigenfuction $Q_p$ determines the mode relaxation
times $\tau_p$, and can be solved numerically, or, in some limits, analytically \cite{ambme1}.
From the bubble-size autocorrelation function
\begin{equation}
A(t)=\langle m(t)m(0)\rangle=\sum_{p\neq 0}A_pe^{-t/\tau_p},
\end{equation}
a typical quantity determined in experiment, we obtain the
relaxation time spectrum $\{\tau_p\}$ with the corresponding amplitudes
$A_p=\left(\sum_{m,n}mQ_p(m,n)\right)^2$. The slowest mode
$\tau_{\mathrm{relax}}=\tau_1$ determines the characteristic
equilibration time \cite{rem1}. These measures can be used to quantify
the bubble-SSB system for different cases:

\emph{(i.) In absence of SSBs},
Eq.~(\ref{master}) reduces to a $(1+1)$-dimensional master equation.
If we neglect the loop closure factor, we can obtain an analytic result
using orthogonal polynomials, from which we infer the
inequalities for the eigenvalues $\tau_p$ \cite{ambme1}:
\begin{equation}
k^{-1}\left(1+u^{1/2}\right)^{-2}\le\tau_M<\ldots<\tau_1\le
k^{-1}\left(1-u^{1/2}\right)^{-2},
\end{equation}
The equal signs hold in the limit
$M\to\infty$. The characteristic relaxation time $\tau_
{\mathrm{relax}}$ tends to very large values (and diverges for $M\to\infty$)
in the limit $u\to 1$, i.e., on approaching $T_m$. In
Fig.~\ref{fig3}, we plot the corresponding relaxation time spectrum for three
temperatures. Well below $T_m$, we see the multistate relaxation observed
experimentally \cite{altan}. For temperatures closer to $T_m$, the slowest
eigenmode becomes increasingly dominant ($\approx 2$-state behavior).
To estimate the rate constant $k$ for zipping we find from Fig.~\ref{fig3} that
at $u=0.6$ $\tau_{\mathrm{relax}}\simeq 20/k$;
comparing to the experimental result $\tau_{\mathrm{relax}}\simeq 10$ms, we
consistently extract $k\simeq 50\mu$s.

\begin{figure}
\includegraphics[width=8cm]{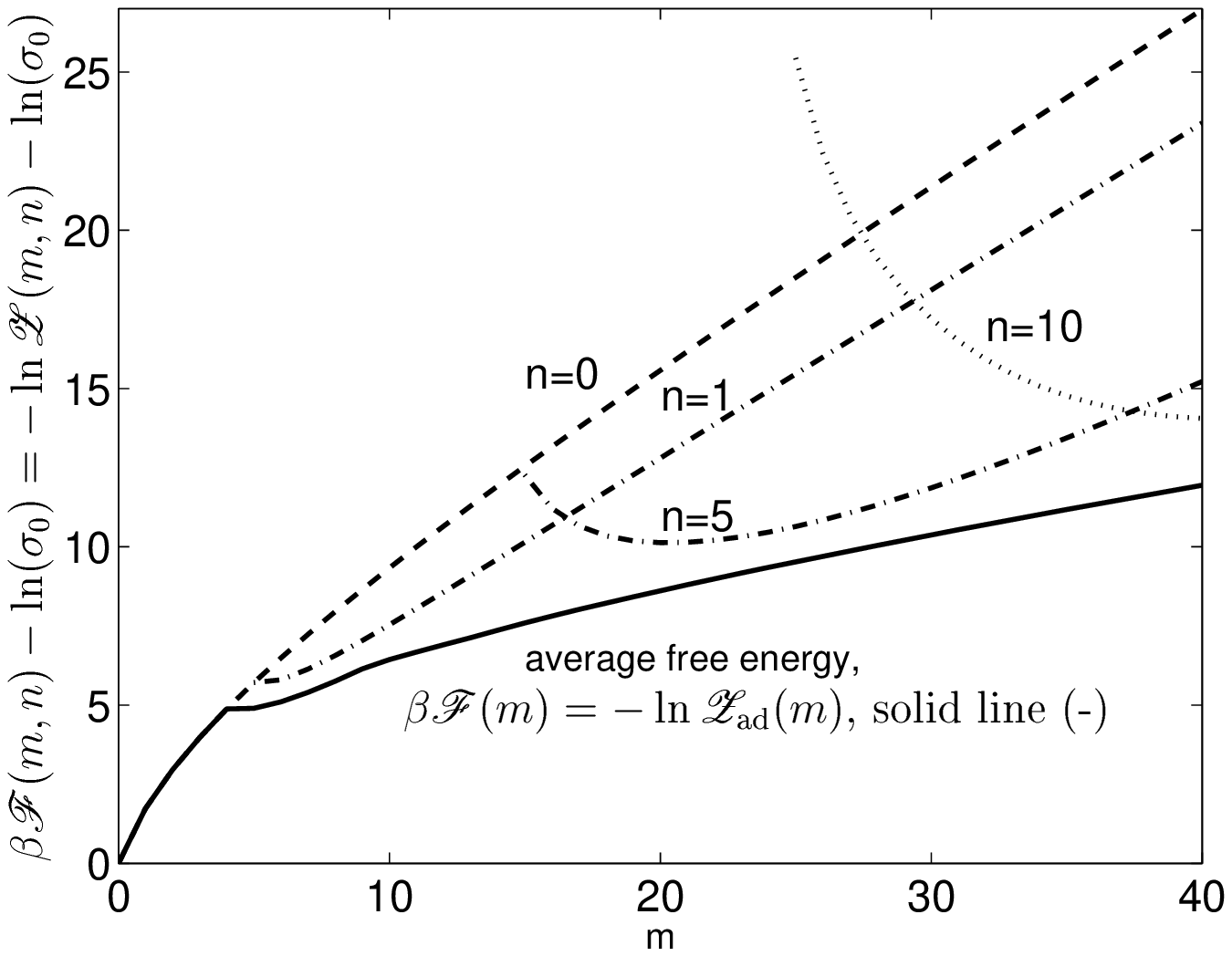}\\
\includegraphics[width=8cm]{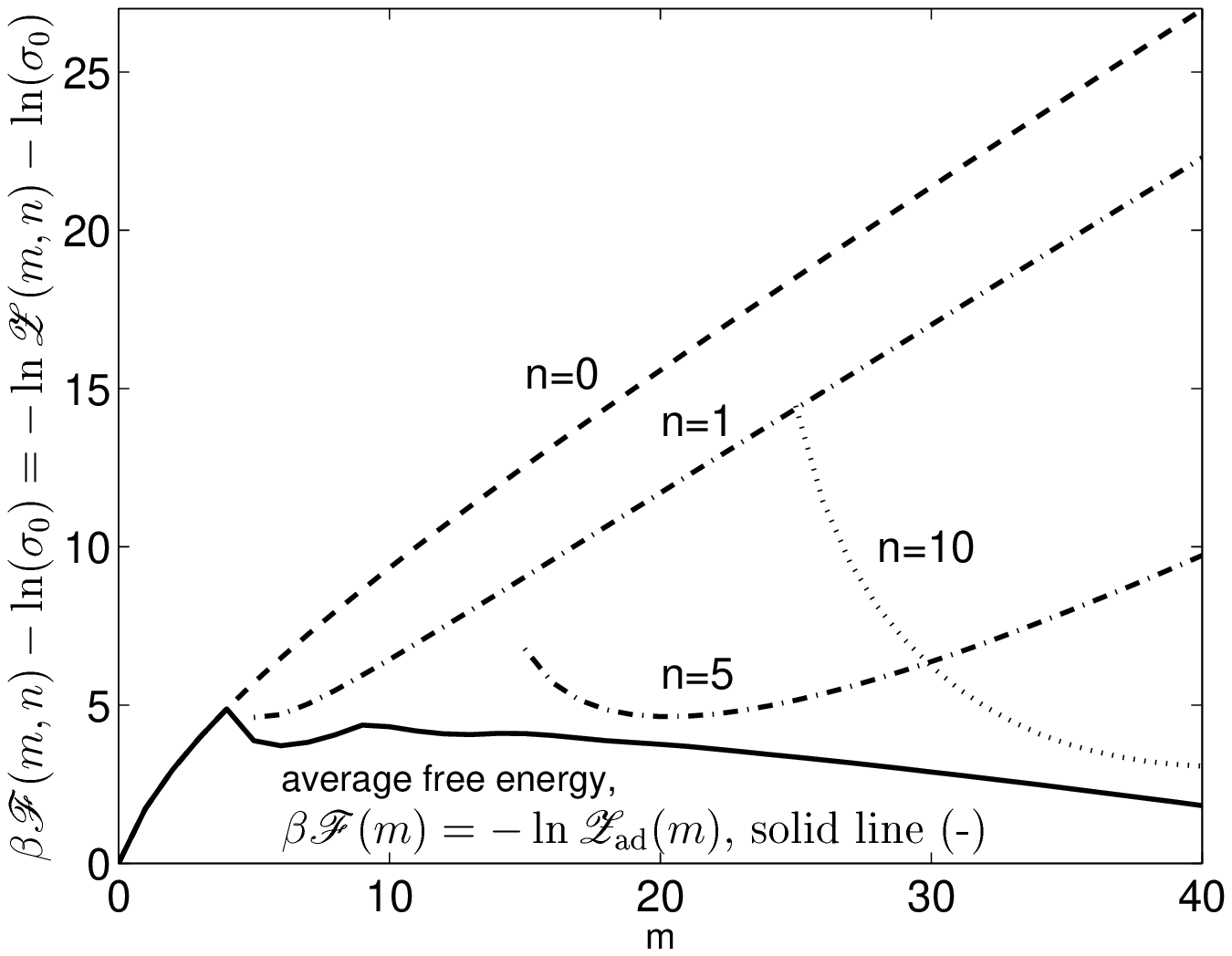}
\caption{Effective free energy in the limit $\gamma\gg 1$ (---),
and `free energy' for various
fixed $n$ ($u=0.6$, $M=40$, $c=1.76$, $\lambda=5$). Top: $\kappa=0.5$; bottom:
stronger binding, $\kappa=1.5$.
\label{fig4}}
\end{figure}

\emph{(ii.) The fast binding limit} corresponds to $\gamma\gg 1$,
i.e., SSB (un)binding being much faster than the bubble zipping.
By adiabatic elimination \cite{risken}, the master equation
(\ref{master}) reduces to an equation in the bubble size $m$
as in case (i.) but where the bubble free energy landscape $\beta\mathscr
{F}=-\ln\mathscr{Z}_{\mathrm{ad}}$, with $\mathscr{Z}_{\mathrm{ad}}=\sum_n
\mathscr{Z}(m,n)$, is dressed by
the ongoing SSB (un)binding. This effective free energy is lowered in
comparison to case (i.), as demonstrated in Fig.~\ref{fig4} for two cases:
(a) the effective free energy is reduced, but the overall slope still positive;
(b) the SSB-dynamics causes a negative slope of the effective free energy,
ultimately leading to complete denaturation of unclamped DNA.
We note a characteristic feature of the effective energy in Fig.~\ref{fig4},
namely, the finite size effects due to the SSB size $\lambda$: the bubble has to
open up to a minimum size $m=\lambda$, before it is able to accommodate the
first SSB, etc. This also effects a nucleation barrier for
SSB-assisted DNA-denaturation even in the case of negative slope of the
dressed free energy for larger $m$. Numerically, we determine the critical
binding strength $\kappa_{\mathrm{crit}}\approx 1$ for the parameters of
Fig.~\ref{fig4}, at which the effective $\beta\mathscr{F}(m)$ is almost flat.
Fast binding is, e.g., realized for gp32 mutants at elevated effective
temperature \cite{pant}.

\begin{figure}
\includegraphics[width=8cm]{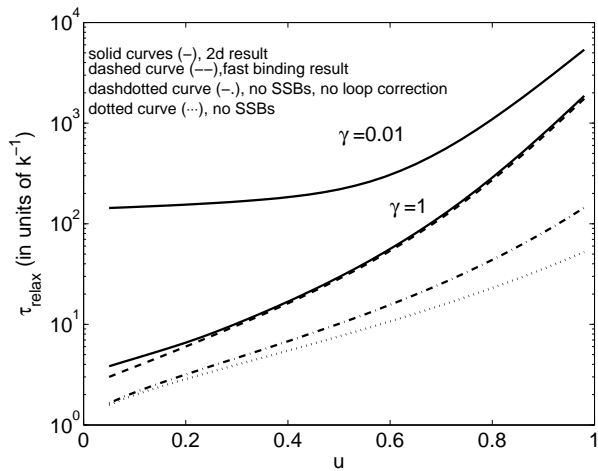}
\caption{Relaxation time as function of $u$, for $\kappa=0.5$,
$\sigma_0=10^{-3}$, $M=20$, $\lambda=5$.
\label{fig5}}
\end{figure}

\emph{(iii.) General case.} If SSB (un)binding is not sufficiently fast but
occurs within the typical bubble relaxation time in absence of SSBs, the
full master equation (\ref{master}) needs to be solved.
By tuning the relative SSB unbinding rate $\gamma$ and SSB binding
strength $\kappa$, we can move from the situation
with practically no binding to the fast binding case. In Fig.~\ref{fig5},
we illustrate this behavior via the characteristic relaxation time $\tau_{
\mathrm{relax}}$.
Experimentally, $u$ and $\kappa$ are changed by variation of temperature and
concentration of SSBs in solution, respectively.

To (un)zip close (open) a base-pair, the two single-strands making up the
bubble have to be pulled closer towards (pushed away from) the zipper fork.
This additional effect may be included using similar
arguments as in Ref.~\cite{dimarzio}. The adjustment of pulling
(pushing) propagates along the contour of the chain until a bend
(inflexion) is
reached, a distance that scales as the gyration radius, i.e., $\simeq m^{\nu}$.
Having in mind Rouse-type dynamics, this would slow down the (un)zipping
rates by a factor $m^{-\nu}$ \cite{ambme1}. The relevance of this effect for
the rather small bubble sizes well below $T_m$ is not obvious, and has to be
based on more accurate experimental or numerical data.

At temperatures well below $T_m$, our description will be valid for unclamped
DNA homopolymers due to $\sigma_0\ll 1$. Explicit boundary and heteropolymer
structure effects can be included by introducing the bubble position as an
additional variable in the master equation (\ref{master}) \cite{hetero}. Our
description will also hold under moderate chain tension, e.g., in combination
with optical tweezers setups. The pulling force $f$ then gives rise to blobs
of size $\xi=k_BT/f$ in which the DNA is undisturbed \cite{degennes}.
Conversely, in the case of strong pulling twist is progressively taken out of
the DNA molecule, corresponding to a torque ${\cal T}$ and the twist energy
$\theta_0{\cal T}$, where $\theta_0=2\pi/10.35$ denotes the twist angle per
base. The statistical weight is modified according to $u\to u\exp(\beta\theta
_0{\cal T})$ \cite{hwa}.

Our (2+1)-dimensional master equation approach provides a
quantitative framework for the coupled dynamics of the size fluctuations of
DNA denaturation bubbles and the binding of SSBs. It explains the different
regimes, from free bubble breathing to SSB-induced denaturation, that
correspond to experimentally accessible in vitro as well as in vivo situations.
In particular, we expect that this scheme is useful to the design of future
experiments, and to estimate in vivo conditions based on the SSB binding
strength $\kappa$, the base stacking factor $u$, and the ratio $\gamma$ of
SSB unbinding and base pair zipping rates.

Given the rather general formulation in terms of a master equation for the
probability distribution $P(m,n,t)$, our approach may be useful to various
other systems where particles bind to a substrate whose binding surface
fluctuates in time (wetting, oxidation or binding of biomolecules to
membranes); in particular, in the presence of finite size effects.
In addition, the coupled DNA bubble/SSB binding
dynamics studied here is a generic example of how a
stochastic process (partially) can rectify another one.

We are happy to thank Richard Karpel, Oleg Krichevsky and Mark Williams for
helpful discussions.


\begin{thebibliography}{99}
\bibitem{REM} The generic behavior of the model does not change substantially
as long as $\sigma_0\ll 1$.
\bibitem{rem} Different choices are possible as detailed balance does not
fully determine the transfer rates.
\bibitem{rem1} In certain cases, it may be of advantage to use the
characteristic relaxation time, and not the longest.
\end{thebibliography}
\end{document}